# Electrically tunable Floquet Weyl photon emission from Dirac semimetal $Cd_3As_2$

**Sobhan Subhra Mishra[1,2], Thomas CaiWei Tan [1,2], Manoj Gupta[1,2], Faxian Xiu[3,4,5], Ranjan Singh[6*]**

[1]*Division of Physics and Applied Physics, School of Physical and Mathematical Sciences, Nanyang Technological University, Singapore 637371*
[2]*Centre for Disruptive Photonic Technologies, Nanyang Technological University, Singapore 639798*
[3]*State Key Laboratory of Surface Physics and Department of Physics Fudan University Shanghai 200433, China*
[4]*Institute for Nanoelectronic Devices and Quantum Computing Fudan University Shanghai 200433, China*
[5]*Shanghai Research Center for Quantum Sciences Shanghai 201315, China*
[6]*Department of Electrical Engineering, University of Notre Dame, Notre Dame, IN, USA*

*\* Corresponding Author- rsingh3@nd.edu*

**Abstract**

The ability to optically engineer the Dirac band and electrically control the Fermi level in two-dimensional (2D) Dirac systems, such as graphene, has significantly advanced quantum technologies. However, similar tunability has remained elusive in three-dimensional (3D) Dirac systems. In this work, we demonstrate both optical and electrical tunability of the band structure in the 3D Dirac semimetal $Cd_3As_2$. Photoexcitation with circularly polarized light breaks time-reversal symmetry, lifting the degeneracy at Dirac points to transform the material into a Floquet Weyl semimetal with chiral Weyl nodes. This transition induces nonzero Berry curvature, giving rise to helicity-dependent transverse anomalous photocurrents, detectable through terahertz (THz) emission at normal incidence. Furthermore, applying an external electric field displaces the Fermi level away from the Dirac point, enlarging the Dirac cone projection leading to a reduced density of states (DoS) of Fermi arcs. As a result, we achieve precise electrical control over Floquet band engineering, resulting in a large modulation of THz emission. Moreover, at oblique incidence, the circular photon-drag effect induces helicity-dependent longitudinal photocurrents. Simultaneous generation and manipulation of both longitudinal and transverse photocurrents enable precise control of the helicity of emitted THz pulses. These results pave the way for electric field-controlled Floquet Weyl THz sources, offering significant potential for applications in quantum computing and low-power electronics.

**Introduction**

Optical and electrical manipulation of the electronic band structure offers immense opportunities to engineer topological properties of topologically trivial systems[1–4]. One notable method involves Floquet band engineering where time-reversal symmetry in 2D Dirac materials, such as graphene, is broken by using circularly polarized (CP) pump light[2,5,6] to create a band gap at the Dirac point and subsequently generating an anomalous Hall photocurrent. The transverse anomalous Hall conductivity $\sigma_y$ can be calculated using the Berry curvature $b_z(\mathbf{k})$, which measures the geometric structure of wave functions in momentum space[7] according to the following equation[2]

$$\sigma_y = \frac{e^2}{\hbar} \int \frac{d^3k}{(2\pi)^3} f(\mathbf{k}) b_z(\mathbf{k}) \qquad (1)$$

Here $f(\mathbf{k})$ is the charge carrier distribution function, $e$ is the electronic charge and $\hbar$ is the reduced Planck's constant.

In a 3D Dirac semimetal such as Cadmium Arsenide ($Cd_3As_2$)[8–12] which serves as a three-dimensional analogue of graphene, the bands do not fully open[6,13] upon breaking the time reversal symmetry via photoexcitation through CP light. Instead, the Dirac nodes split into pairs of doubly degenerate Weyl nodes, transforming from a Dirac semimetal into a Weyl semimetal[14] state known as a Floquet Weyl Semimetal[6,15], as illustrated Figure 1(e)[1,6,14]. Its non-trivial topology induces a significant Berry curvature around the nodes, thus engendering a helicity dependent transverse anomalous photocurrent[1,16,17], as described by Equation 1 and consequently a helicity dependent terahertz (THz) radiation.

While prior research has extensively investigated the optical control of Berry curvature and helicity-dependent phenomena in topological materials[1,2,6,16], the role of electrical control in these processes remains largely unexplored. Introducing an external electric field modifies the carrier distribution $f(\mathbf{k})$ shown in Equation 1, providing an opportunity to actively tune the anomalous photocurrents. This capability not only enables precise modulation of THz emission but also offers a new dimension of control that addresses challenges in electrically tunable THz source design[18].

Additionally, when illuminated with circularly polarized light at an oblique angle, the angular momentum transfer between the incident photons and the photoexcited carriers produces a helicity-dependent longitudinal photocurrent through circular photon drag effect (CPDE). Consequently, simultaneous generation and manipulation of longitudinal and electrically tunable transverse THz waves highlight the potential for creating an electric field controlled chiral THz source[19,20], a limitation that persists despite extensive research into various THz emission processes such as optical rectification in various nonlinear crystals like *ZnTe*[21], spin[22], and orbital dynamics[23,24] of magnetic heterostructures. Moreover, the generated helicity dependent transverse photocurrent is directly proportional to Berry curvature[2,6,13], hence giving a direct experimental measure of the quantized topological charges in the Weyl nodes.[25,26]

In this study, we demonstrate THz emission from $Cd_3As_2$ thin films at various pump polarization which was controlled by changing the quarter wave plate (QWP) angle. Using a phenomenological equation, we have decoupled the helicity-independent thermal and shift current contributions from the helicity-dependent contributions to the THz emission. By distinguishing these mechanisms, we establish that under circularly polarized pump condition at normal incidence, the THz emission from $Cd_3As_2$ predominantly arises through Floquet band engineering. The application of an external electric field allows further control of the magnitude of the transverse photocurrent with 30% and 60% modulation in the emitted THz when pumped with linear and circularly polarized light respectively. Moreover, at oblique incidence, a helicity dependent longitudinal photocurrent is also generated due to circular photon drag effect. As a result, a precise control of polarization of the emitted THz wave is achieved with demonstration of tuning of ellipticity from 5% to 98% covering half of the Poincaré sphere. Our approach enables the advancement of electrically tunable Floquet Weyl chiral THz photon sources, paving the way for design of energy efficient THz source.

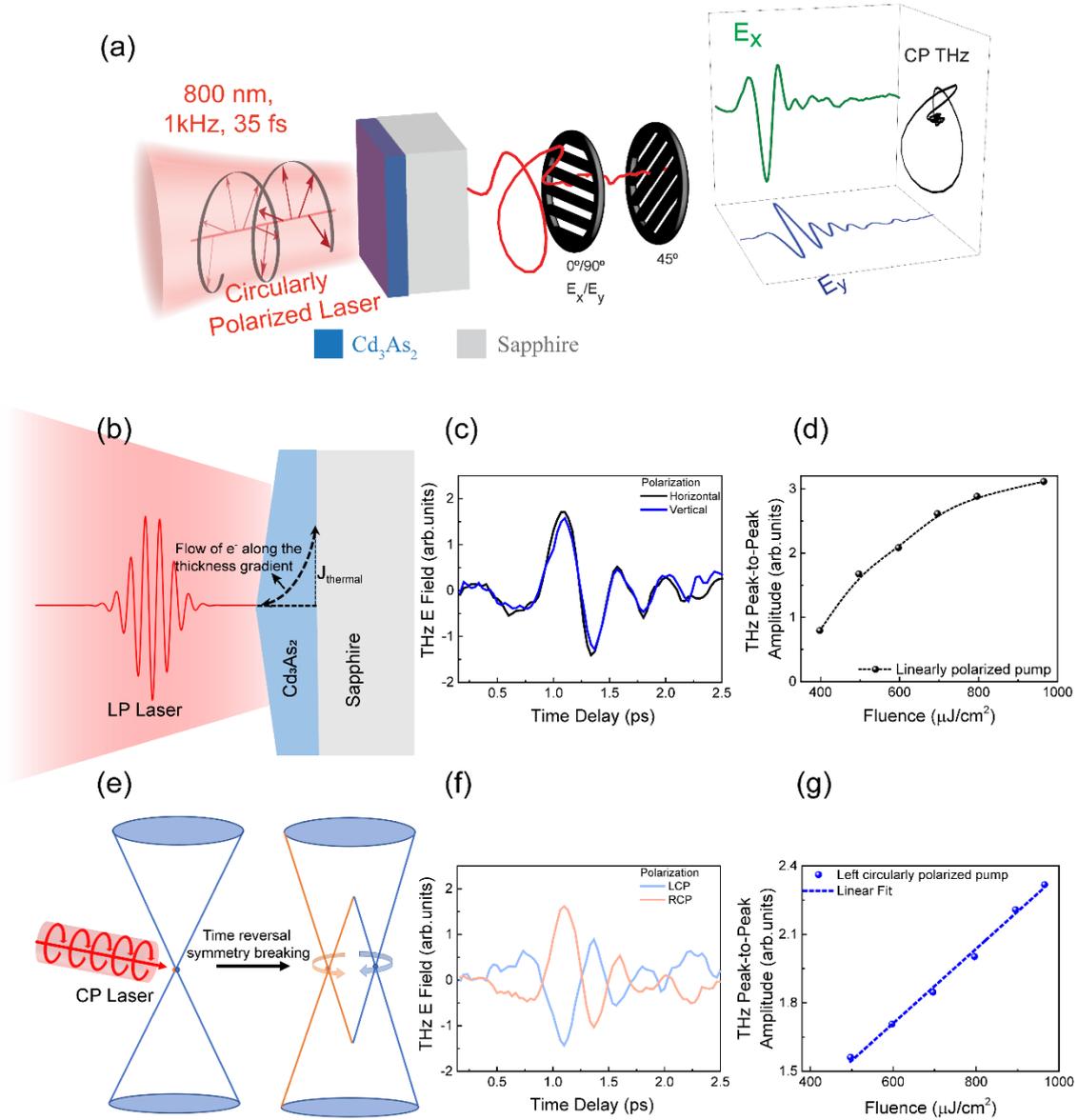

**Figure 1: Terahertz emission from $Cd_3As_2$ when excited with linearly and circularly polarized light.** (a) Schematic of Chiral THz emission set up with $Cd_3As_2$ as an emitter (b) Mechanism of THz emission due to ultrafast photothermal effect due to the thickness gradient in $Cd_3As_2$ thinfilm when pumped with linearly polarized light; (c) Polarization independent THz emission when pumped with linearly polarized light; (d) Nonlinear rise in the peak-to-peak amplitude of the emitted THz pulse with increasing fluence suggesting the presence of ultrafast photothermal effect when pumped with linear polarized light; (e) Mechanism of THz emission due to Floquet band engineering when pumped with circularly polarized light; (f) Helicity dependent THz emission when pumped with circularly polarized light; (g) Linear rise in the peak-to-peak amplitude of the emitted THz pulse with increasing fluence indicating a second-order nonlinear process when pumped with left circularly polarized light;

**Results and Discussion**

A 40 nm $Cd_3As_2$ thin film was grown on a 420 μm C-cut sapphire substrate via Molecular Beam Epitaxy, in presence of a 20 nm *CdTe* buffer layer. The sample was then irradiated with an 800 nm laser beam operating at 1 kHz repetition rate and a 35-fs pulse width at a normal incident angle. A detailed figure of the setup is shown in supplementary section S1, and chiral THz emission schematic is shown in Figure 1(a). Figure 1(b) illustrates the mechanism of THz emission from $Cd_3As_2$ under linearly polarized pump conditions. When the pump is normally incident, the radial thickness gradient of the film induces an in-plane temperature gradient, leading to THz emission[28]. This emission is independent of the polarization of the pump laser, as evidenced by identical THz emission profiles with both horizontally and vertically polarized light, as shown in Figure 1(c). Notably, the peak-to-peak amplitude of the emitted THz waveform displays a nonlinear dependence on increasing pump fluence, as depicted in Figure 1(d), distinguishing it from typical second-order nonlinear effects. However, under circularly polarized pump conditions, due to Floquet band engineering, a pronounced helicity dependent THz pulse is emitted. Figure 1(e) shows the schematic of the formation of Floquet Weyl semimetal state when $Cd_3As_2$ is pumped with circularly polarized light. Emitted THz waves for left and right circularly polarized photoexcitation are depicted in Figure 1(f). In this case, when pumped with left circularly polarized light, the phase of the emitted THz is opposite in comparison to the phase when pumped with linear and right circularly polarized light, indicating a generation of negative photocurrent. Additionally, the THz pulse amplitude increases linearly with pump fluence as shown in Figure 1(g) when pumped with left circularly polarized light, supporting the presence of a second-order nonlinear process, and ruling out photothermal effects.

To investigate the underlying mechanism further, the peak-to-peak amplitude of the emitted signal was measured as a function of pump polarization, which was varied by adjusting the Quarter Wave Plate (QWP) angle. The polarization dependence of the amplitude of the emitted THz pulse, displayed in Figure 2(a), was fitted to the phenomenological model described below in equation 2[27,28]. This fitting enabled the extraction of coefficients representing the distinct contributions to THz emission from $Cd_3As_2$, as illustrated in Figure 2(b)

$$E_{THz} = C sin2\alpha + L_1 sin4\alpha + L_2 cos4\alpha + D \qquad (2)$$

Here the first term ($C sin2\alpha$) explains the helicity and polarization dependent contribution towards THz emission, the second ($L_1 sin4\alpha$) and third ($L_2 cos4\alpha$) terms capture the polarization dependent but helicity independent THz emission, and the last term ($D$) describes the helicity and polarization independent THz emission due to ultrafast photothermal effect[29]. $L_1$ describes the shift currents which arise due to a spatial shift in the charge carrier position resulting in an interband optical excitation generating a time-dependent dipole moment. Since these currents are symmetry-constrained, they require an electric field component normal to the surface. Therefore, at normal incident angle, their contribution towards THz emission is negligible as shown in Figure 2(b). Moreover, the shift current contribution vanishes at both linear and circular polarized photoexcitation. $L_2$ and $D$ depict the thermal effects associated with linear absorption with $L_2$ describing the current due to linear photon drag effect and $D$ accounting for the ultrafast photothermal effect. Quantitatively, the helicity dependent term is about twice of the of the helicity independent term and is of negative value as shown in Figure 2(b) indicating that the THz emission is predominantly from helicity dependent mechanism when time reversal symmetry is broken. Additionally, it was also observed that $L_2$ and $D$ have comparable values.

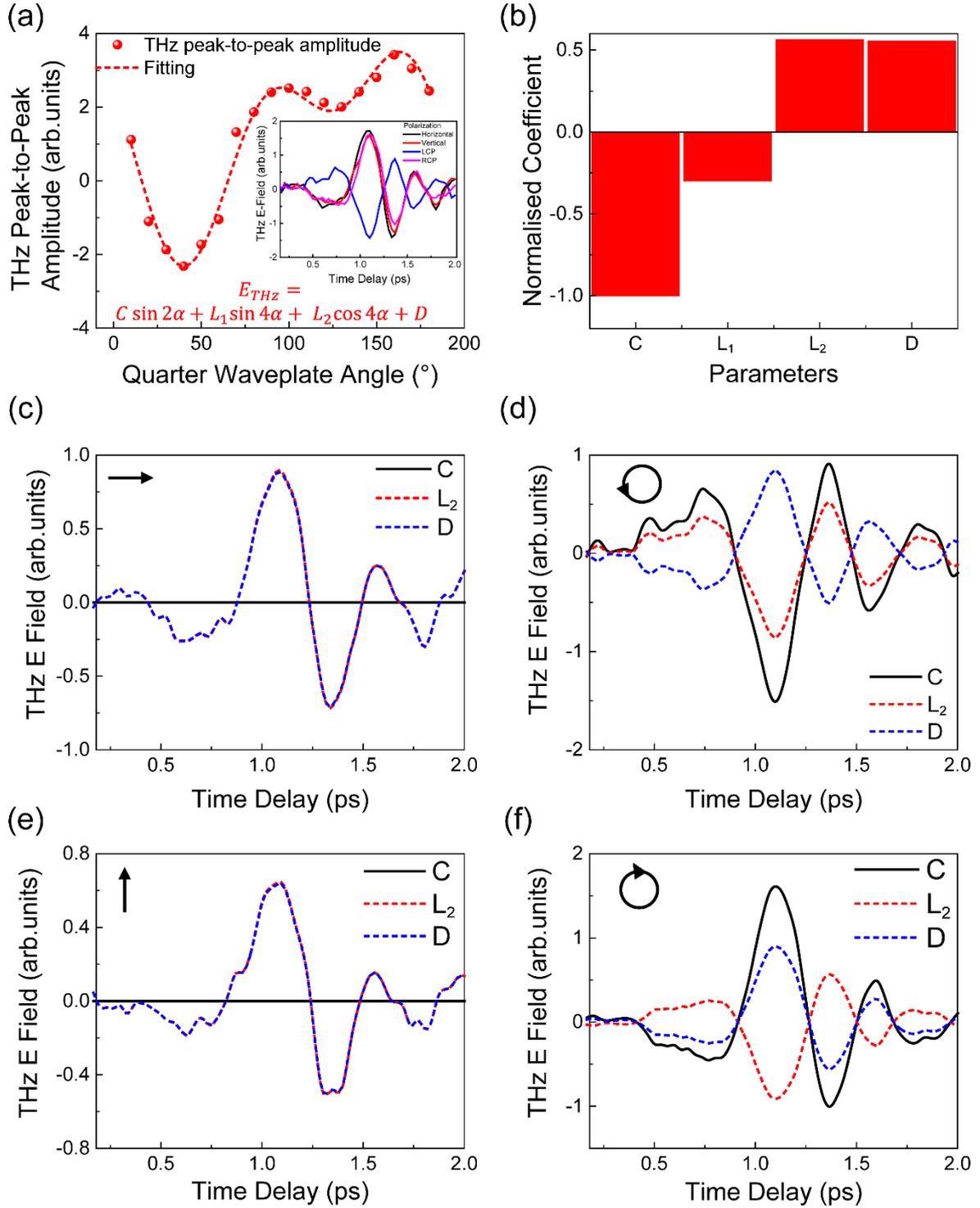

**Figure 2: Disentanglement of different contribution and reconstructed THz signal due to various factors** (a) Extracted THz peak-to-peak amplitude at different QWP angle ($\alpha$) for optical pump light. Dotted line is the fitted curve according to the phenomenological equation 2, inset shows the THz pulses at both the linear and circularly polarized pumps; (b) Normalized extracted fitting parameters attributed to the factors affecting the emitted THz; Reconstructed THz pulse due to each factor when pump polarization is (c) Horizontal (d) Left circular (e) Vertical (f) Right circular. When the pump beam is circularly polarized, $L_2$ and $D$ cancel each other resulting in THz emission mainly due to helicity dependent term (Floquet band engineering)

To gain further understanding of the system, the contributions of the THz pulse from individual components described in Equation 2 were reconstructed and presented for both linear and circular polarized pumps in Figure 2(c-f). For these polarizations, THz pulse contribution from the second term ($L_1 sin4\alpha$) will be zero since $sin4\alpha$ equals zero when $\alpha$ = 0°, 45°, 90°, or 135°. Under linear pump polarization ($\alpha$ = 0°, 90°), the helicity-dependent term ($C sin2\alpha$) also vanishes, but the remaining two terms ($L_2 cos4\alpha$ and $D$) add constructively to produce a stronger THz pulse, as illustrated in Figures 2(c) and 2(e). When the pump is left circularly polarized in nature ($\alpha$ = 45°), the helicity-dependent term ($C sin2\alpha$) becomes negative due to the negative value of $C$, resulting in phase reversal in the emitted THz pulse, as shown in Figure 2(d). Notably, the other two terms ($L_2 cos4\alpha$ and $D$) cancel each other out, as $L_2$ and $D$ are of comparable magnitude (see Figure 2(b)), making the resultant THz pulse entirely attributed to the helicity- dependent term. When the pump polarization is right-circular ($\alpha$ = 135°), the phase of the THz pulse is opposite to the emitted THz when the pump is left circular ($\alpha$ = 45°) in nature as shown in Figure 1(e), and the helicity-dependent component of the THz pulse is similarly inverted, as depicted in Figure 2(d) and 2(f).

When a 3D Dirac semimetal like $Cd_3As_2$ having linearly dispersing energy bands is photoexcited with a circularly polarized light, due to Floquet band engineering, a transverse anomalous photocurrent is generated. The resultant anomalous hall conductivity when the pump polarization is left circular can be given by[13,30]

$$\sigma_y = - \frac{N_D e^4 v E_{pump}^2}{2\pi^2 \hbar^3 \omega^3} \qquad (3)$$

Here, $N_D$ represents the number of Dirac node, $v$ denotes the Fermi velocity, $E_{pump}$ is the amplitude of the pump pulse, and $\omega$ signifies the frequency of the pump pulse. The negative sign indicates that the anomalous Hall conductivity, induced by Floquet band engineering, becomes negative when subjected to left circularly polarized pumping[13]. The negative anomalous hall conductivity can be attributed to negative Chern number of -1 of the nonzero Berry curvature induced by the left circular polarized photoexcitation. The detailed derivation of the anomalous hall photoconductivity can be found in supplementary section S4[13,30]. Therefore, photocurrent and consequently THz generated, flip its phase when pumped with left circular light as depicted in Figure 1(f) and 2(d).

To achieve electric-field tuning of THz emission from $Cd_3As_2$, an out-of-plane capacitive device was engineered. Aluminum (Al) electrodes (100 nm) were deposited on the edges of the $Cd_3As_2$ thin film, serving as one of the metal plates of the capacitor. Silicon dioxide ($SiO_2$, 1 $\mu$m thick) was deposited as the dielectric material, followed by a THz transparent Indium Tin Oxide (ITO) coated Polyethylene terephthalate film (PET) substrate layer (surface resistivity of 300 Ω/sq) as the second metal plate. The capacitance of the structure was measured to be 1.15 nF over an overlap area $0.8\ cm \times 0.8\ cm$ of two electrodes. An out-of-plane electric field was applied to $Cd_3As_2$, with the device schematic shown in Figure 3(a). As previously discussed, under linearly polarized light, THz emission is primarily governed by an ultrafast photothermal effect due to the thickness gradient in the $Cd_3As_2$ thin film. However, applying a positive electric field causes the electrons to drift directly towards the ITO electrode rather than moving along the thickness gradient towards the center, resulting in a reduction of thermal photocurrent, as seen in Figure 3(b). This effect leads to a gradual decrease in THz emission with increasing positive voltage under linear pump polarization, as demonstrated in Figure 3(d). A clear modulation in THz emission of approximately 30% is achieved, as illustrated in Figure 3(f). Furthermore, applying an out-of-plane electric field drives the Fermi level away from the Dirac point, as illustrated in Figure 3(c). This displacement enlarges the bulk Dirac cone projection, leading to a reduced density of states (DOS) of the Fermi arcs and consequently diminishing the photocurrent, which reduces the THz emission. Figure 3(e) shows this decrease in THz pulse emission under left circularly polarized light. Notably, a modulation of approximately 60% is achieved at a 90V bias when pumped with circularly polarized light, as shown in Figure 3(f). The effect can be further amplified by using thinner films of $Cd_3As_2$. To confirm the electrical tuning of THz emission through Floquet band engineering, THz peak-to-peak amplitude under varying polarization and bias conditions were extracted and fitted using the phenomenological Equation 2, as displayed in Figure 3(g). The distinct components contributing to THz emission were then obtained and plotted against different voltages in Figure 3(h), revealing a clear reduction in the helicity-dependent component $C$ with increasing voltage and a corresponding increase in the helicity-independent contribution $D$ with $L_1$ and $L_2$ remaining largely constant. Figures 3(b) and 3(d) confirm that the photothermal current decreases with the application of a positive voltage suggesting that the observed

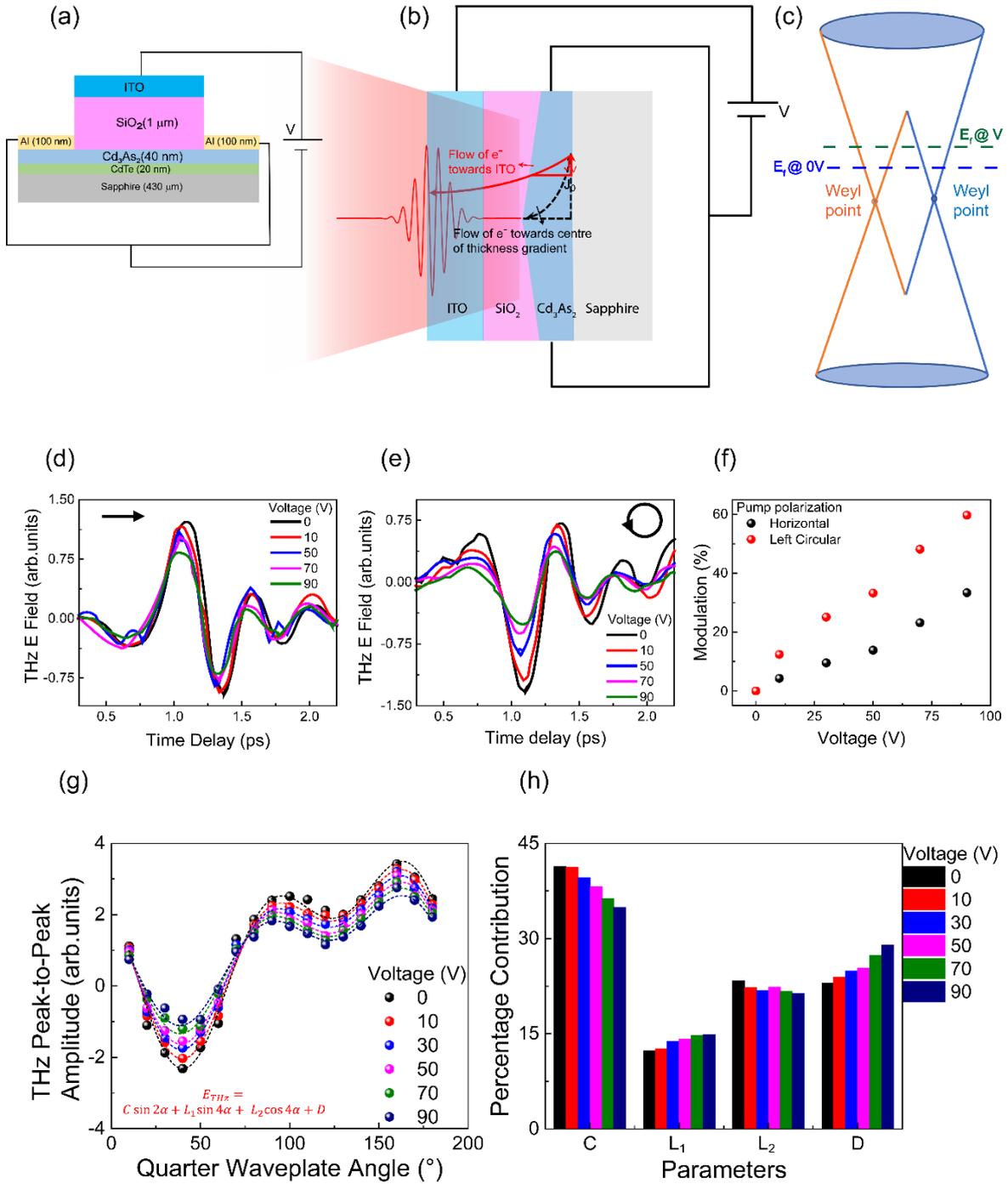

**Figure 3: Electrical tuning of the Anomalous photocurrent.** (a) Device schematic of electrical tuning of $Cd_3As_2$; (b) Mechanism of E field tunable photothermal effect when pumped with linear polarized light; (c) Mechanism of E field tunable fermi level when pumped with circularly polarized light; (d) Electrical tuning of the emitted THz when pumped with linear polarized light; (e) Electrical tuning of THz emission when pumped with circular polarized light; (f) Comparison of electrical modulation of the emitted pulse when pumped with linear and circular polarized light; (g) Extracted THz amplitude at different QWP angle ($\alpha$) at different applied voltage. Dotted line is the fitted curve according to the phenomenological Equation 2; (h) Electrical tuning of the normalized extracted fitting parameters attributed to the factors affecting THz emission

increase in $D$ with applied voltage is a consequence of the reduction in $C$, rather than the opposite.

Finally, to explore the possibility of emission of Chiral THz from $Cd_3As_2$, the transverse and longitudinal THz components were recorded independently using two wire grid polarizers as depicted in Figure 1(a) and Supplementary section S1. A 130 nm pristine $Cd_3As_2$ is taken to increase the Signal to Noise ratio while capturing both the polarization. As shown in Figure 4(b), when illuminated by circularly polarized light at normal incidence, the longitudinal component of the emitted THz pulse is almost negligible and it increases as the incident angle is increased, signifying a pronounced angular dependence. In contrast, the transverse component, which arises from Floquet band engineering, exhibits a significant amplitude even at normal incidence as demonstrated in Figure 4(a). This phenomenon can be attributed to the tilting of the dispersion relation along $k_z$ direction in $Cd_3As_2$ because of crystallographic anisotropy, resulting in a split not only in momentum but also in energy, thus generating a non-zero photocurrent. A slight change in the amplitude was observed with increase in incident angle which can be possible due to possible presence of out of plane spin texture of the Dirac cone at oblique incident. Since both transverse and longitudinal photocurrents are simultaneously generated under circularly polarized light, the emitted THz pulse is capable of supporting both Y and X polarized THz emission at small incident angles. As shown clearly in Figure 4(c), when $Cd_3As_2$ is photoexcited with left circular polarized light, phase of the Y polarized emitted THz signal, generated from transverse photocurrent, flips, reflecting the reversal in the direction of the photocurrent. Additionally, phase of the Y polarized emitted THz under right circular pump condition is opposite to the phase of the Y polarized emitted THz under left circular pump condition. This demonstrates that the helicity dependent Y polarized THz pulse originates due to transverse photocurrent from Floquet band engineering. Moreover, due to circular photon drag effect, a helicity dependent longitudinal photocurrent is also generated resulting in the generation of X polarized THz emission as shown in Figure 4(d). There is a clear temporal shift in the X polarized emitted THz pulse with respect to Y polarized emitted THz pulse when pumped with circularly polarized light due to different relaxation dynamics of the charge carriers participating in transverse and longitudinal photocurrents[31] resulting in a possible phase difference. Figure 4(e) shows the phase difference between X and Y polarized emitted THz pulse when pumped with LCP and RCP light at an incident angle of 10°. The phase of the emitted THz is obtained by performing the fast Fourier transform of the recorded time pulses. As indicated, the phase difference between X and Y polarized emitted THz is $\frac{\pi}{2}$ for RCP and $\frac{-\pi}{2}$ for LCP at 10° incident angle. Moreover, the amplitude of both the polarization is almost equal giving rise to the chiral THz photons. The projection of X and Y polarized THz pulse by plotting the amplitudes from the time pulses can be seen in Supplementary section S5 indicating a circular THz wave generation when the pump polarization is circular in nature. However, when pumped with linearly polarized light, the time reversal symmetry in the band is preserved resulting in a linear Dirac band structure. As a result, the emitted THz is always dominated by the thermal effects, thus generating a linear vertically polarized THz wave. A clear illustration of polarization control is shown on the Poincaré sphere in Figure 4(f). Each point on the sphere represents a distinct polarization state of the emitted THz photons. The Poincaré sphere is characterized by the azimuth angle ψ (latitude 2ψ) and the ellipticity χ (longitude 2χ). Linear polarization is represented by points along the equator. As the points move away from the equator latitudinally, the degree of circular polarization increases. Points at the poles indicate circular polarization, and other elliptical polarization states are represented by points elsewhere on the sphere except the equator. For a unit sphere, the Cartesian coordinates of these points are related to the spherical coordinates by the equation: $x = cos(2\chi)cos(2\psi), y = cos(2\chi)sin(2\psi)$ and $z = sin(2\psi)$ for $0 \leq \psi < \pi$ and $-\frac{\pi}{4} < \chi < \frac{\pi}{4}$. As a result, the ellipticity angle, related to χ, governs the cartesian coordinate of the z-axis to quantify the degree of circular polarization, where $z = 1$ is obtained for right circularly polarized wave, and $z = -1$ is obtained for left circularly polarized wave. The details of the construction of Poincaré sphere are given in methods section. In Figure 4(f), the reconfigurable polarization states of the emitted THz radiation are demonstrated across eight different points, corresponding to eight distinct pump ellipticities. The polarization states of the emitted pulse are tuned from left circularly polarized (χ = -41.85°) when pumped with left circularly polarized optical light, to vertically linear (χ = 2.25°) when pumped with linearly polarized optical light, and to right circularly polarized (χ = 44.1°) when pumped with right circularly polarized optical light indicating a complete latitudinal control of THz polarization and chirality.

In summary, by breaking the time reversal symmetry in the Dirac semimetal cadmium arsenide ($Cd_3As_2$) using circularly polarized pump light, transverse helicity dependent photocurrent is generated due to the Floquet band engineering, resulting in emission of chiral Weyl photons which is further controlled by application of an external electric field. Additionally, at oblique incidence, transfer of angular momentum from incident photon to excited charge carriers facilitates the emission of helicity dependent longitudinal THz photons due to circular photon drag effect. Simultaneous generation of longitudinal and electrically tunable transverse photocurrent allows for the generation of electrically tunable chiral THz photons. Our work highlights the potential of Floquet band engineering as platforms for generating electrically tunable and optically driven THz emitter. The ability to break time-reversal symmetry and electrically engineer chiral THz emission paves the way for designing novel THz devices with applications in ultrafast wireless communication, advanced spectroscopy, and quantum information systems. Expanding this approach to other non-equilibrium condensed matter state could unlock new quantum phases and broaden the functionality of next-generation photonic and optoelectronic technologies.

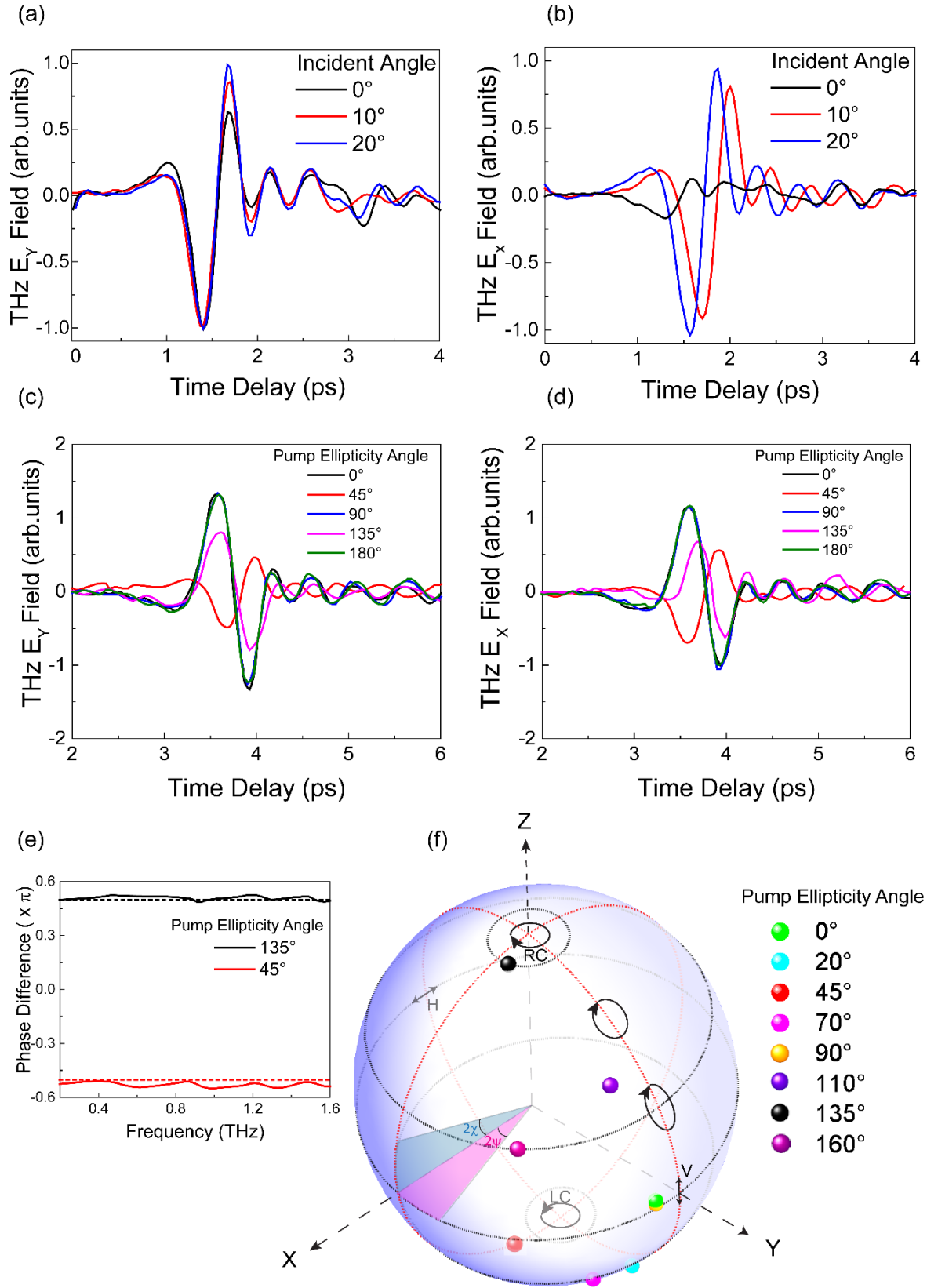

**Figure 4: Emission and control of Chiral THz from $Cd_3As_2$** (a) Transverse component of the emitted THz pulse when pump polarization is left circular at different incident angle indicating negligible dependence of the THz with incident angle; (b) Longitudinal component of the emitted THz pulse when pump polarization is left circular at different incident angle indicating negligible THz at normal incidence

and clear increase in the emitted pulse with increase in incident; (c) Y (d) X Polarized THz emitted from $Cd_3As_2$ at different pump polarization; (e) phase difference between E$_X$ and E$_Y$ at circular polarization; (f) Poincaré sphere depicting eight different measured polarization states of the emitted THz pulse at various pump polarization.

**Methods**

***Sample Preparation:*** The $Cd_3As_2$ thin film was deposited using a molecular beam epitaxy system (PerkinElmer 425B, Waltham, MA). A c-cut sapphire substrate of thickness 420 μm with <0001> orientation was used. It was annealed at 550° C for 30 mins to thoroughly remove the unwanted absorbed molecules. A 20 nm *CdTe* buffer layer was deposited to reduce the mismatch between the $Cd_3As_2$ film and substrate. A high purity (99.999%) $Cd_3As_2$ source material from dual filament and valve cracker effusion cells was evaporated onto the CdTe layer to grow the film epitaxially at 120° C. The thickness was monitored by in-situ reflection high energy electron diffraction (RHEED) system.

***THz emission experiment***: The emitted THz radiation is captured using a 1 mm thick *ZnTe* crystal oriented along the <110> axis. The femtosecond laser pulse, illuminating the $Cd_3As_2$ thin film has a wavelength of 800 nm, corresponding to a laser energy of 1.55 electron volts (eV). It operates at a repetition rate of 1 kHz with a pulse width of 35 femtoseconds (fs). A beam splitter divides the laser pulse into two parts: the higher intensity laser beam is directed towards the photoexcitation of the emitter, while the lower intensity laser beam serves as a probe for detection. A quarter waveplate (QWP) is employed at the pump side to change the polarization of the beam used to photoexcite the sample. A mechanical delay stage is used for precise time matching of the THz and probe beam at the detector crystal. The emitted THz pulse is collected by 4 parabolic mirrors and focused on the *ZnTe* <110> detector, inducing birefringence in the crystal. Simultaneously, the time-matched probe beam traverses the crystal and experiences a change in polarization directly proportional to the birefringence. For electro-optic detection, a quarter-wave plate and a Wollaston prism are used to distinguish the *s* and *p* polarizations of the laser. The rotational changes in the probe laser are subsequently detected using a balanced photodiode that measures the intensity difference between the *s* and *p* polarized light. The electrical signal thus detected is initially pre-amplified before being input into a lock-in amplifier to enhance the signal-to-noise ratio. The resultant signal from the lock-in amplifier is used to generate the THz electric field. Additionally, as illustrated in Figure 1 and Supplementary section S1, a combination of two wire-grid polarizers (WGP) is employed to separately record the $E_Y$ and $E_X$ components of the THz pulse. WGP-1 is utilized to distinguish the $E_Y$ and $E_X$ components of the field, while WGP-2 is consistently set at a 45° angle to transmit the orthogonal THz component with the same output polarization, ensuring uniform detection at the *ZnTe* detector. The entire set up is kept under Nitrogen environment to minimize the humidity of the surroundings. A detailed set up picture is provided in supplementary section S1.

### *Construction of the Poincaré sphere:*

To have a better visualization of the polarization state of the emitted THz pulse a Poincaré sphere was constructed, as shown in Figure 4(f). In a unit sphere, the cartesian coordinates and the spherical coordinates are related via the following equations:

$$x = \cos(2\chi)\cos(2\psi), 0 \leq \psi < \pi$$

$$y = \cos(2\chi)\sin(2\psi), -\frac{\pi}{4} < \chi < \frac{\pi}{4}$$

$$z = \sin(2\chi)$$

where, $\psi$ and $\chi$ are the spherical coordinates and $x^2 + y^2 + z^2 = 1$ is the equation of the unit sphere.

The spherical coordinates indicated by $\psi$ and $\chi$ of the Poincaré sphere are defined as the parameters of the polarization ellipse (azimuth angle: $\psi$ and ellipticity angle: $\chi$) from the following equations.

$$\tan 2\psi = \frac{2 E_Y E_X}{E_X^2 - E_Y^2} \cos \delta \qquad 0 \leq \psi \leq \pi$$

$$\sin 2\chi = \frac{2 E_X E_Y}{E_X^2 + E_Y^2} \sin \delta \qquad -\frac{\pi}{4} \leq \chi \leq \frac{\pi}{4}$$

where, $\delta$ is the phase difference between the X and Y polarized light,

$E_X$ is the electric field amplitude of X-polarized light

$E_Y$ is the electric field amplitude of Y-polarized light

The phase difference $\delta$ for all the pump polarizations are calculated from the Fast Fourier Transform (FFT) of the emitted THz time pulses, as shown in main text Figure 4(e). The electric field amplitude for X and Y polarized light $E_X$ and $E_Y$ are recorded separately by employing a combination of 2 wire-grid polarizers, as shown in Figure 1(a).

**Data availability**

The data supporting this study's findings are available from the corresponding author upon reasonable request. Supplementary information is linked to the online version of the paper.

**Acknowledgments**

S.M. thanks Dr. Baolong Zhang and Nikhil Navaratna for valuable discussions and suggestions. S.M, T.C.T, M.G and R.S acknowledge funding support from National Research Foundation (NRF) Singapore, Grant nos. NRF-CRP23-2019-0005 (TERACOMM) and NRF-MSG-2023-0002 (NCAIP). F.X. was supported by the National Natural Science Foundation of China (52225207, 11934005, and 52350001), the Shanghai Pilot Program for Basic Research - FuDan University 21TQ1400100 (21TQ006).

**Competing Interests**

The authors declare no competing interests.

**References**

1. Sato, S. A. *et al.* Light-induced anomalous Hall effect in massless Dirac fermion systems and topological insulators with dissipation. *New J. Phys.* **21**, 093005 (2019).
2. McIver, J. W. *et al.* Light-induced anomalous Hall effect in graphene. *Nat. Phys.* **16**, 38–41 (2020).
3. Wang, Y. H., Steinberg, H., Jarillo-Herrero, P. & Gedik, N. Observation of Floquet-Bloch States on the Surface of a Topological Insulator. **342**, (2013).
4. Rudner, M. S. & Lindner, N. H. Band structure engineering and non-equilibrium dynamics in Floquet topological insulators. *Nat Rev Phys* **2**, 229–244 (2020).
5. Broers, L. & Mathey, L. Observing light-induced Floquet band gaps in the longitudinal conductivity of graphene. *Commun Phys* **4**, 248 (2021).
6. Oka, T. & Kitamura, S. Floquet Engineering of Quantum Materials. *Annu. Rev. Condens. Matter Phys.* **10**, 387–408 (2019).
7. Berry Michael Victor 1984 "Quantal phase factors accompanying adiabatic changesProc. R. Soc. Lond. A39245–57"
8. Wang, Z., Weng, H., Wu, Q., Dai, X. & Fang, Z. Three-dimensional Dirac semimetal and quantum transport in $Cd_3As_2$. *Phys. Rev. B* **88**, 125427 (2013).
9. Neupane, M. *et al.* Observation of a three-dimensional topological Dirac semimetal phase in high-mobility Cd3As2. *Nat Commun* **5**, 3786 (2014).
10. Liang, T. *et al.* Ultrahigh mobility and giant magnetoresistance in the Dirac semimetal Cd3As2. *Nature Mater* **14**, 280–284 (2015).
11. Crassee, I., Sankar, R., Lee, W.-L., Akrap, A. & Orlita, M. 3D Dirac semimetal Cd3As2: A review of material properties. *Phys. Rev. Materials* **2**, 120302 (2018).
12. Dai, Z. *et al.* High Mobility 3D Dirac Semimetal ($Cd_3As_2$) for Ultrafast Photoactive Terahertz Photonics. *Adv. Funct. Mater.* **31**, 2011011 (2021).
13. Murotani, Y. *et al.* Disentangling the Competing Mechanisms of Light-Induced Anomalous Hall Conductivity in Three-Dimensional Dirac Semimetal. *Phys. Rev. Lett.* **131**, 096901 (2023).
14. Armitage, N. P., Mele, E. J. & Vishwanath, A. Weyl and Dirac semimetals in three-dimensional solids. *Rev. Mod. Phys.* **90**, 015001 (2018).
15. Hübener, H., Sentef, M. A., De Giovannini, U., Kemper, A. F. & Rubio, A. Creating stable Floquet–Weyl semimetals by laser-driving of 3D Dirac materials. *Nat Commun* **8**, 13940 (2017).
16. Bucciantini, L., Roy, S., Kitamura, S. & Oka, T. Emergent Weyl nodes and Fermi arcs in a Floquet Weyl semimetal. *Phys. Rev. B* **96**, 041126 (2017).
17. Weyl spin-momentum locking in a chiral topological semimetal | Nature Communications. https://www.nature.com/articles/s41467-024-47976-0.
18. Agarwal, P., Huang, L., Ter Lim, S. & Singh, R. Electric-field control of nonlinear THz spintronic emitters. *Nat Commun* **13**, 4072 (2022).
19. Agarwal, P. *et al.* Reconfigurable Chiral Spintronic THz Emitters. *Advanced Optical Materials* **12**, 2303128 (2024).
20. Gao, Y. *et al.* Chiral terahertz wave emission from the Weyl semimetal TaAs. *Nat Commun* **11**, 720 (2020).
21. Vidal, S., Degert, J., Tondusson, M., Freysz, E. & Oberlé, J. Optimized terahertz generation via optical rectification in ZnTe crystals. *J. Opt. Soc. Am. B* **31**, 149 (2014).
22. Seifert, T. *et al.* Efficient metallic spintronic emitters of ultrabroadband terahertz radiation. *Nature Photon* **10**, 483–488 (2016).

23. Mishra, S. S., Lourembam, J., Lin, D. J. X. & Singh, R. Active ballistic orbital transport in Ni/Pt heterostructure. *Nat Commun* **15**, 4568 (2024).
24. Seifert, T. S. *et al.* Time-domain observation of ballistic orbital-angular-momentum currents with giant relaxation length in tungsten. *Nat. Nanotechnol.* (2023) doi:10.1038/s41565-023-01470-8.
25. De Juan, F., Grushin, A. G., Morimoto, T. & Moore, J. E. Quantized circular photogalvanic effect in Weyl semimetals. *Nat Commun* **8**, 15995 (2017).
26. Ji, Z. *et al.* Spatially dispersive circular photogalvanic effect in a Weyl semimetal. *Nat. Mater.* **18**, 955–962 (2019).
27. Cheng, L. *et al.* Giant photon momentum locked THz emission in a centrosymmetric Dirac semimetal. *Sci. Adv.* **9**, eadd7856 (2023).
28. Chen, Z. *et al.* Defect-induced helicity dependent terahertz emission in Dirac semimetal PtTe2 thin films. *Nat Commun* **15**, 2605 (2024).
29. Lu, W. *et al.* Ultrafast photothermoelectric effect in Dirac semimetallic Cd3As2 revealed by terahertz emission. *Nat Commun* **13**, 1623 (2022).
30. Zhang, X.-X., Ong, T. T. & Nagaosa, N. Theory of photoinduced Floquet Weyl semimetal phases. *Phys. Rev. B* **94**, 235137 (2016).
31. Murotani, Y. *et al.* Anomalous Hall Transport by Optically Injected Isospin Degree of Freedom in Dirac Semimetal Thin Film. *Nano Lett.* **24**, 222–228 (2024).

# Supplementary Information

## Section S1: Terahertz emission spectroscopy setup

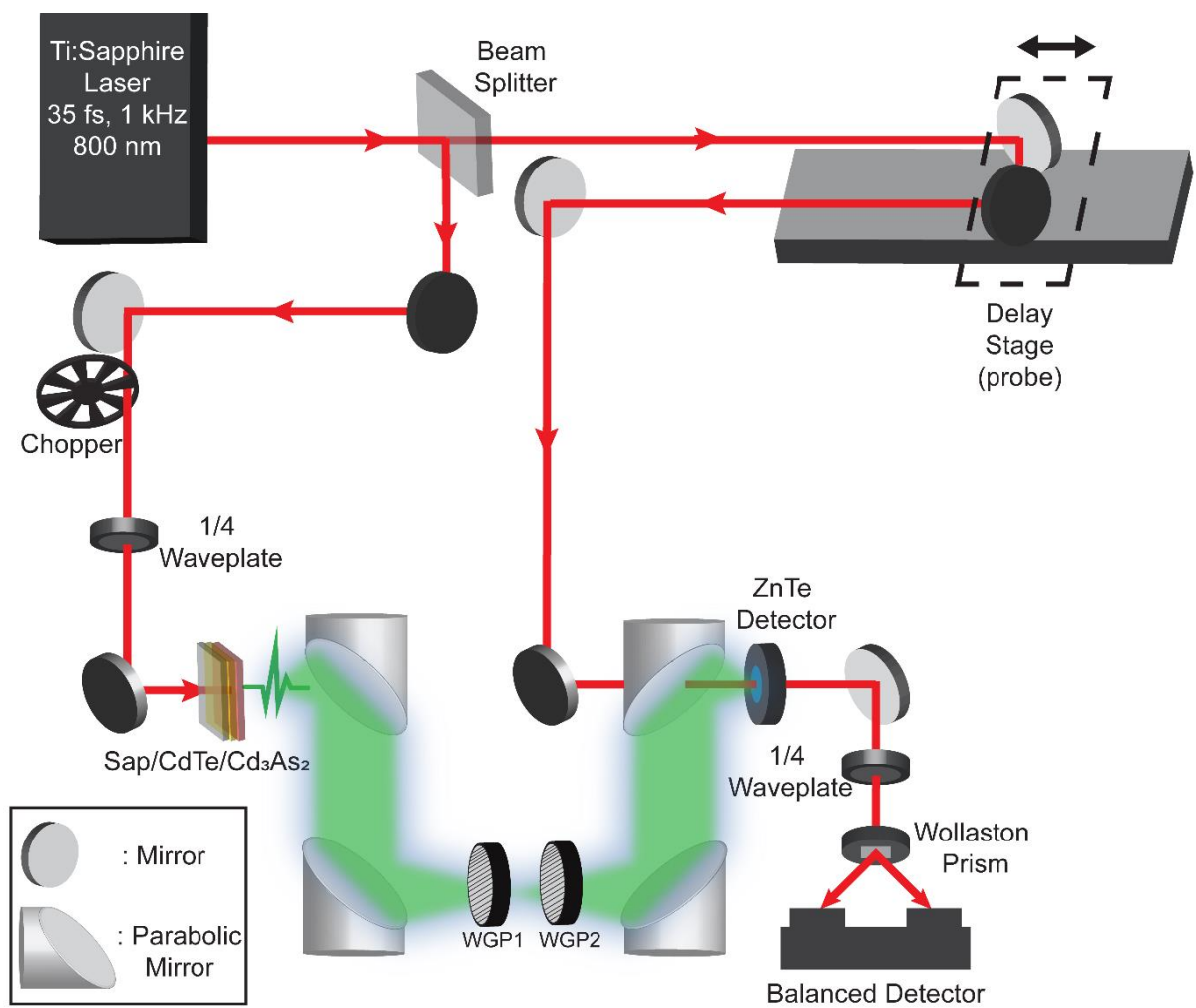

**Figure S1: THz emission spectroscopy set up for chiral THz generation from $Cd_3As_2$**

Figure S1 shows the Chiral THz emission spectroscopy setup with $Cd_3As_2$ as a Terahertz emitter. A quarter waveplate (QWP) is introduced in the pump path to photoexcite $Cd_3As_2$ with different polarized light. 2 wire grid polarizer was employed before the detector to separate the transverse and longitudinal component of the emitted THz. The 2nd QWP is always kept at 45° and the 1st QWP is kept at 0°/90° to extract $E_x$ and $E_y$ respectively.

**Section S2: Terahertz transmission spectroscopy of $Cd_3As_2$**

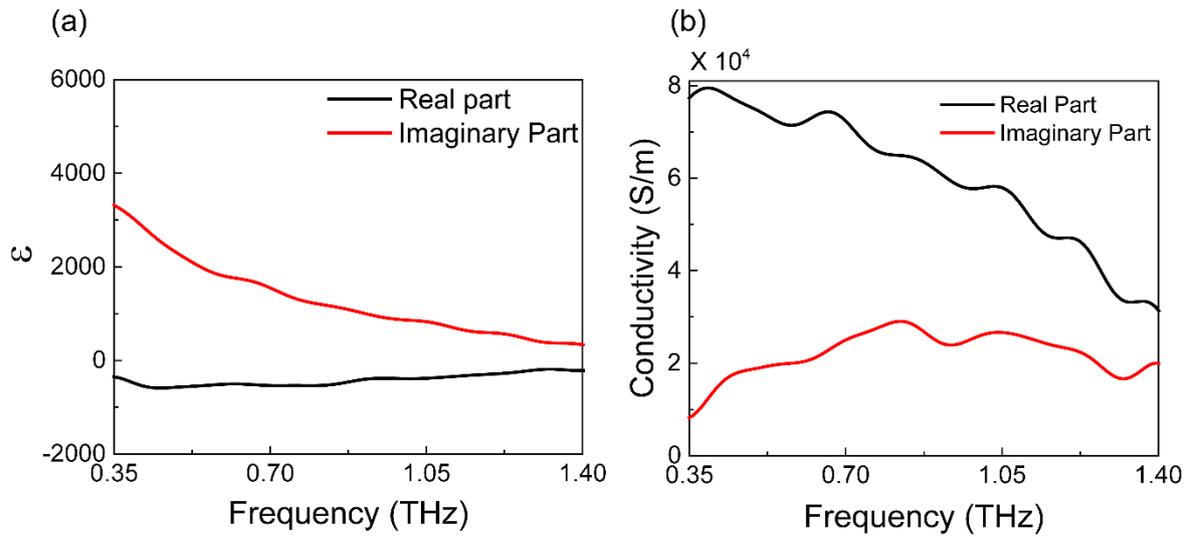

**Figure S2: THz transmission spectroscopy for 130 nm $Cd_3As_2$ thinfilm** (a) Real and imaginary part of (a) dielectric constant of $Cd_3As_2$ thinfilm (b) complex conductivity of $Cd_3As_2$ thin film.

Figure S2 shows the permittivity and conductivity of 130 nm $Cd_3As_2$ measured using THz time domain spectroscopy. The conductivity could be fitted by Drude Smith expression to calculate the scattering time of the thin film which was calculated to be around 160 fs indicating the good quality of the film[1].

## Section S3: Comparison of THz generation from $Cd_3As_2$ to *ZnTe*

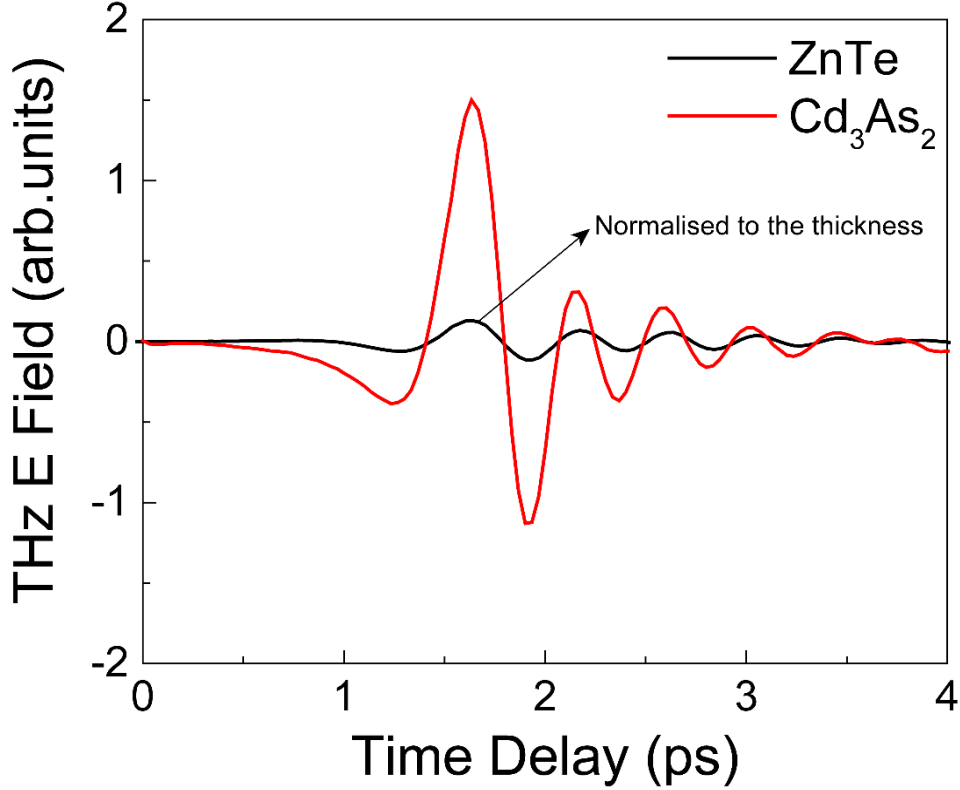

**Figure S3: Normalized THz emission from $Cd_3As_2$ and *ZnTe***

Figure S3 shows the THz emission normalized to thickness and pump fluence for *ZnTe* and $Cd_3As_2$. It is clear that $Cd_3As_2$ emits THz at around 10 times more efficiently than *ZnTe*.

## Section S4: Photoinduced Floquet Weyl semimetal

Let us assume that the Dirac node is isotropic in nature. The Dirac node consists of two degenerate chiral Weyl nodes for which the Hamiltonian can be written in the following way

$$H_R(\mathbf{k}) = \hbar v \mathbf{k}.\boldsymbol{\sigma}, \quad H_L(\mathbf{k}) = -\hbar v \mathbf{k}.\boldsymbol{\sigma} \quad (1)$$

Here $H_R$ and $H_L$ represent the right and left-handed chirality of the degenerate Weyl nodes and $v$ represents Fermi velocity. Pauli matrices $\boldsymbol{\sigma}$ **is** related to orbital angular momentum of the wave function. The linear dispersion relation for the degenerate left and right-handed nodes can be calculated to be

$$\epsilon_\pm(\mathbf{k}) = \pm\hbar v k \qquad (2)$$

And berry curvatures $b_{R\pm}$ and $b_{L\pm}$ are given by

$$b_{R\pm}(\mathbf{k}) = -b_{L\pm}(\mathbf{k}) = \mp\frac{k}{2k^3} \qquad (3)$$

For an ideal Dirac semimetal, the right-handed Weyl node acts as a source, and the left-handed Weyl node acts as a sink for the Berry curvature.

When coupled to an external field, for an instance an LCP light, the system can be modified by replacing $\mathbf{k}$ by $\mathbf{k} - \Delta\mathbf{k}$ where $\Delta\mathbf{k} = \frac{eA(t)}{\hbar}$

Here $e$ is the electronic charge, and $\mathbf{A}(t)$ is the vector potential of the perturbation field. For a left circular polarized light driving field

$$\mathbf{A}(t) = \frac{E_{pump}}{\omega}(\cos\omega t, \sin\omega t, 0)$$

Here $E_{pump}$ is the pump field amplitude which can be calculated from the pump fluence and the pump pulse width.

The Floquet band can then be written as Floquet Hamiltonian[2]

$$H^{nm} = \frac{\omega}{2\pi}\int_0^{\frac{2\pi}{\omega}} (e^{i(n-m)\omega t}H(t) - \delta^{nm}n\hbar\omega)\,dt$$

For an incident energy $\hbar\omega \gg$ the energy scale of interest, the integral can be expanded for high frequency. The resultant Hamiltonian is

$$H^{eff} = H^{00} + \sum_{n=1}^{\infty} \frac{1}{n\hbar\omega}[H^{0n}, H^{0(-n)}]$$

Therefore, the effective Hamiltonian for the individual Weyl nodes under LCP would be

$$H_R^{eff} = \hbar v\,(\mathbf{k} - \Delta\mathbf{k}).\boldsymbol{\sigma} \text{ and } H_L^{eff} = -\hbar v\,(\mathbf{k} + \Delta\mathbf{k}).\boldsymbol{\sigma} \qquad (4)$$

where

$$\Delta \boldsymbol{k} = \frac{e^2 v E_{pump}^2}{\hbar^2 \omega^3} (0,0,1)$$

From equation 4 it can be concluded that, if photoexcited with circular polarized light, Weyl nodes in the Dirac semimetal are not degenerate anymore in momentum space, thus a Floquet Weyl semimetal state emerges. From the left- and right-handed effective Hamiltonian we can see that the right-handed Weyl nodes move in $+k_z$ direction whereas the left-handed Weyl nodes move in $-k_z$ direction.

If the Fermi energy lies at these Weyl nodes, the Chern number which can be defined as the integration of three-dimensional Berry curvature texture can be calculated by integrating the following equation through the Brillouin zone as

$$C(k_z) = \int \frac{dk_x dk_y}{2\pi} b_n^z(\boldsymbol{k}) \qquad (5)$$

which is equal to -1 between the two Floquet Weyl nodes. Therefore, the anomalous Hall conductivity becomes negative as shown in equation 3 in the main text.

**Section S5: Projection of $E_X$ and $E_Y$**

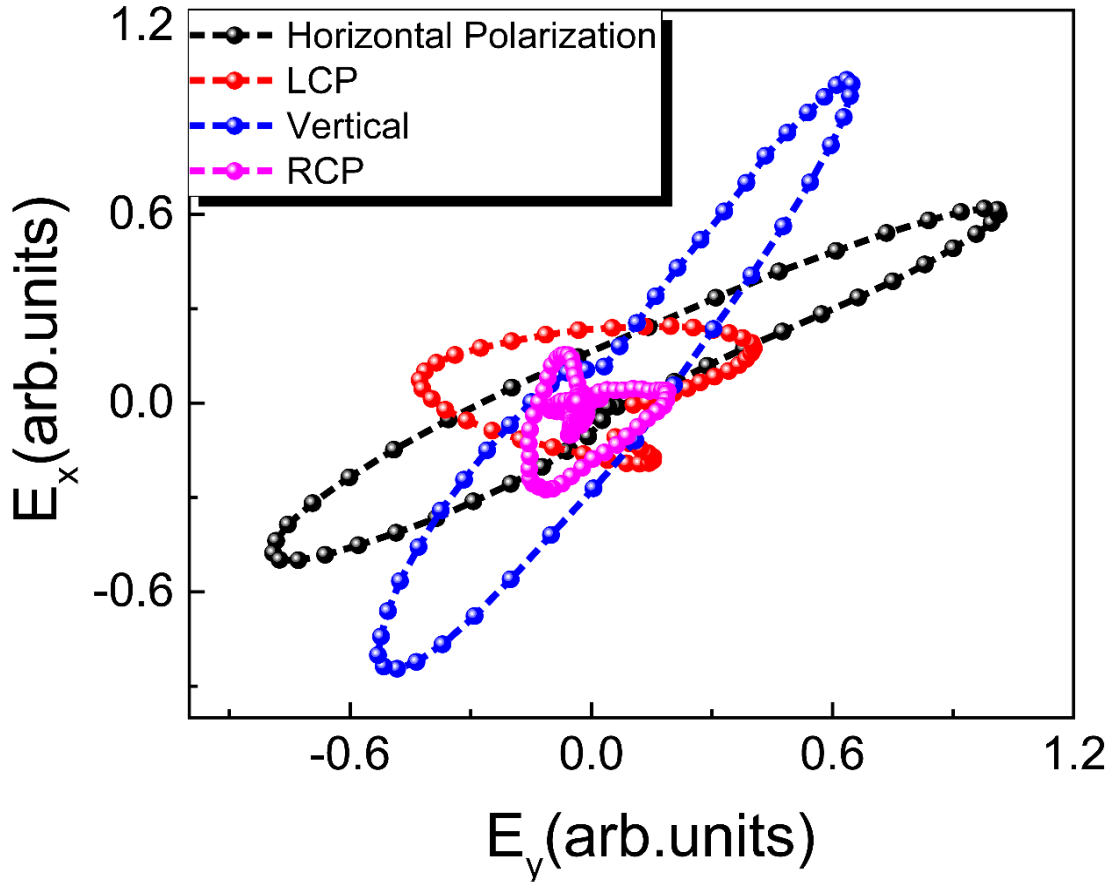

**Figure S4: Generation of circularly polarized terahertz waves when the polarization of the pump laser pulses is changed from linear to circular**

Figure S4 indicates the projection of X and Y polarized THz pulse by plotting the amplitudes from the time pulses. The projection shows a circular THz wave generation when the pump polarization is circular in nature. However, when pumped with linearly polarized light the time reversal symmetry in the band is preserved resulting in a linear Dirac band structure. As a result, the emitted THz is always dominated by the photothermal effect and linear photon drag effect, thus generating a vertically linear THz wave.

**References**

1. Dai, Z. *et al.* High Mobility 3D Dirac Semimetal ($Cd_3As_2$) for Ultrafast Photoactive Terahertz Photonics. *Adv. Funct. Mater.* **31**, 2011011 (2021).
2. Zhang, X.-X., Ong, T. T. & Nagaosa, N. Theory of photoinduced Floquet Weyl semimetal phases. *Phys. Rev. B* **94**, 235137 (2016).